\documentstyle[epsfig]{aipproc}
\newcommand{\beq}{\begin{equation}}
\newcommand{\eeq}{\end{equation}}
\newcommand{\bea}{\begin{eqnarray}}
\newcommand{\eea}{\end{eqnarray}}
\newcommand{\ket}[1]{\left| #1\right>}

\newcommand{\prl}{{\it Phys. Rev. Lett.} }
\newcommand{\rem}[1]{ }
\begin{document}
\title{Self-Interacting Dark Matter with Flavor Mixing}

\author{Mikhail V. Medvedev }
\address{Canadian Institute for Theoretical Astrophysics,
University of Toronto,\\ Toronto, Ontario, M5S 3H8, Canada}

\maketitle

\begin{abstract}
The crisis of the cold dark matter  and 
problems of the self-interacting dark matter  models 
is resolved by postulating flavor mixing of dark matter particles.
Flavor-mixed particles segregate in the gravitational field to 
form dark halos composed of heavy mass eigenstates. Since
these particles are mixed in the interaction basis, elastic 
collisions convert some of heavy eigenstates into light ones 
which leave dense central regions of the halo. This annihilation-like 
process will soften dense central cusps of halos. The proposed model 
accumulates most of the attractive features of self-interacting and 
annihilating dark matter models, but does not suffer from their severe 
drawbacks. This model is natural; it does not require fine tuning. 
\end{abstract}

\section*{Introduction}

Dark matter constitutes most of the mass in the Universe, but its nature
and properties remain largely unknown. A model of structure formation in a 
universe with cold dark matter (CDM) is in excellent agreement with 
observations on large scales ($\gg$~Mpc) which, thus, supports the hypothesis 
that dark matter particles are heavy and weakly interacting with baryonic 
matter and photons, while on small (galactic and sub-galactic scales) it 
appears to be in conflict with recent observations. This fact suggests that 
the CDM model is a good first approximation, but it has to be corrected on 
a galactic scale. 

The simplest and most popular models are the self-interacting dark matter
(SIDM) \cite{SS00} and the annihilating dark matter (ADM) \cite{KNT00}.
There are some problems with each. In a simplest SIDM, the scattering 
cross-section of dark matter (DM) particles, $\sigma_{si}$, must be such 
that to ensure that the flat core forms just by now; larger $\sigma_{si}$
result in core collapse and cuspy cores while for smaller $\sigma_{si}$
the core flattening time is larger than the Hubble time. Moreover, it
seems difficult to obtain flattened cores in galaxies and in clusters 
by the same time because of very different dynamical scales.
The ADM model suffers from the severe ``annihilation catastrophe''
in the early universe (unless some {\it ad hoc} assumptions made).
Here we present a natural model which has all attractive features of 
these both models but does not suffer from their drawbacks.

\section*{SIDM model with Quantum Mixing}

For the sake of simplicity, let us assume that there exist two flavors,
the strongly self-interacting, $\ket{\cal I}$, and the non-self-interacting
or ``sterile'', $\ket{\cal S}$, ones. Each of these interaction eigenstates is
a superposition of two mass eigenstates, the ``heavy'', $\ket{M}$, 
and the ``light'', $\ket{m}$, ones. We write 
\begin{equation}
{\ket{M}\choose{\ket{m}}}=
\left(\begin{array}{cc}
\cos\vartheta & \sin\vartheta\\
-\sin\vartheta & \cos\vartheta
\end{array}\right)
{\ket{\cal I}\choose{\ket{\cal S}}},
\label{mix}
\end{equation}
where $\vartheta$ is the mixing angle. Throughout the paper, we assume for 
simplicity that $M\gg m$, so that heavy states are {\em non-relativistic} 
and light states are {\em relativistic}. 

\begin{figure}[b!] 
\centerline{\epsfig{file=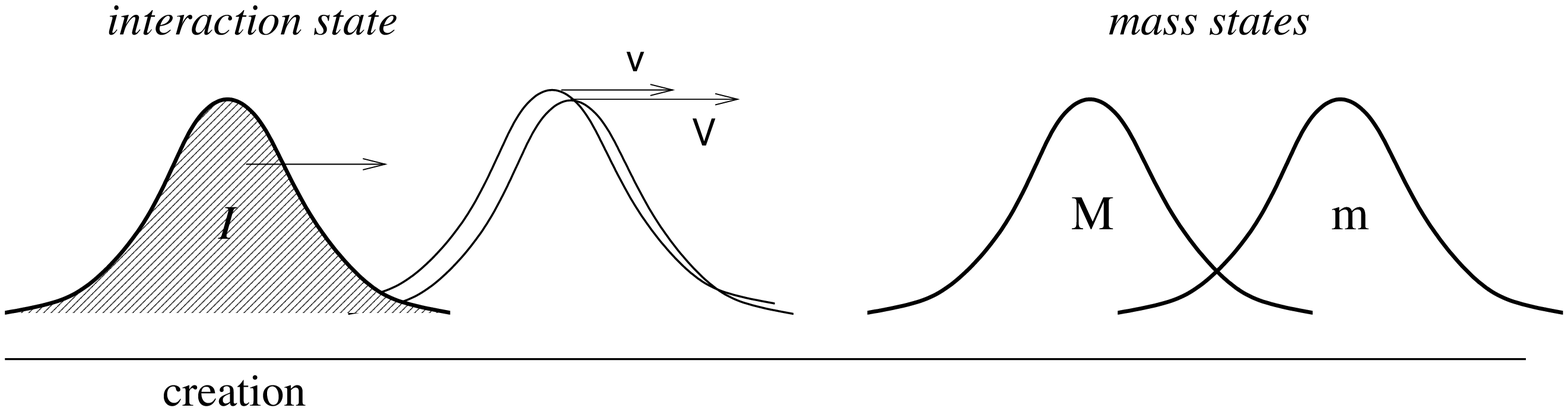,width=3.6in}}~\\
\centerline{\epsfig{file=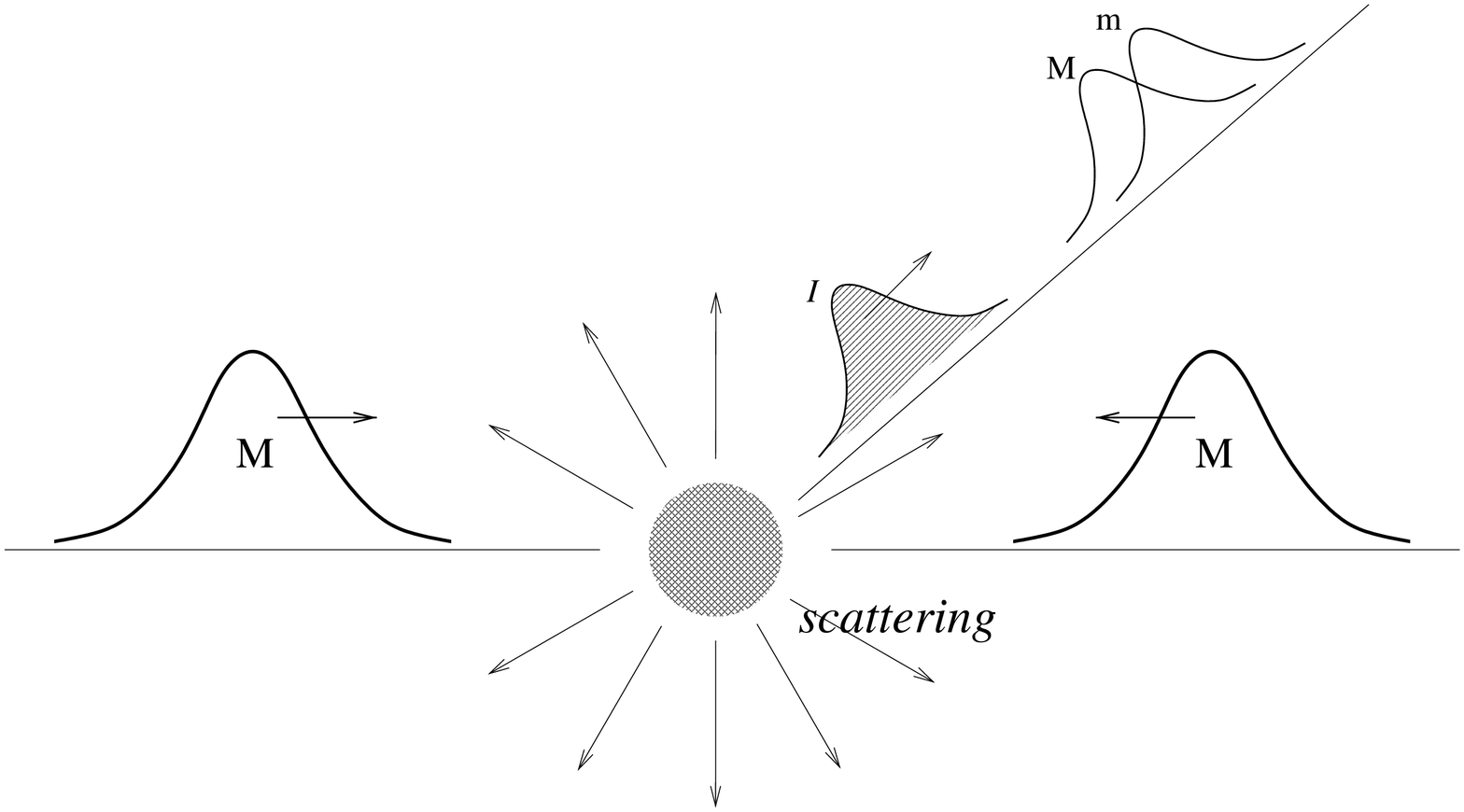,width=3.6in}}
\vspace{10pt}
\caption{Illustration of separation of mass states ({\it a})
and of the scattering of two heavy states which after all
leads to formation of light states ({\it b}).}
\label{fig1}
\end{figure}

The concept of flavor eigenstates arises when interactions of particles
are considered. In the field-free theory the particle fields in the mass
basis have a physical meaning instead. In general, such mass eigenstates have 
different momenta and energies, $E^2_{M,m}=|{\bf p}_{M,m}|^2c^2+\{M,m\}^2c^4$. 
Between the interactions, the mass eigenstates propagate independently, with 
different velocities ${\bf v}_{M,m}={\bf p}_{M,m}c^2/E_{M,m}$. Thus at times
\begin{equation}
t\gg t_s\sim\delta x/|{\bf v}_m-{\bf v}_M|\sim\delta x/c,
\end{equation}
where $\delta x$ is the spatial width of a wave-packet, these states are 
separated from each other and their wave functions no longer overlap, as 
illustrated in Fig.\ \ref{fig1}a. The separation time above is negligibly small
compared to a galactic dynamical time scale. In a gravitational field 
different eigenstates also segregate by mass. Non-relativistic $\ket{M}$-states 
form halos and the large-scale structure (this is CDM) while relativistic 
$\ket{m}$-states leave the halo and behave similar to hot DM.

As dark matter halos form, the density in the central parts increases
and, at some point, self-interactions of the dark matter particles become
important. Let us consider the elementary act of {\em elastic} scattering
of two DM particles, i.e., $\ket{M}$ eigenstates.  The initial wave
function of two interacting particles in the center of mass frame, 
according to equation (\ref{mix}), is
\begin{equation}
\Psi_i=\left(e^{ikz}\pm e^{-ikz}\right)\ket{M}
=e^{\pm ikz}\left(\cos\vartheta\ket{\cal I}+\sin\vartheta\ket{\cal S}\right),
\end{equation}
where ``$+$'' corresponds to $\Psi_i$ symmetric to interchange of particles 
(integer total spin) and ``$-$'' -- to an antisymmetric $\Psi_i$ (half-integer 
total spin), the exponents represent two waves, propagating to the right 
and to the left, and $e^{\pm ikz}\equiv\left(e^{ikz}\pm e^{-ikz}\right)$ 
is the short-hand notation. For scattering, the interaction basis is 
appropriate, rather than the mass basis, hence the expansion above. During 
the scattering event, only $\ket{\cal I}$-component is changing, since 
$\ket{\cal S}$-component does not interact. 
The wave function after scattering at large distances  thus becomes
\begin{eqnarray}
\Psi_s&\approx&\left(e^{\pm ikz}+\frac{f_\pm(\theta)}{r}\,e^{ikr}\right)
\cos\vartheta\ket{\cal I}+e^{\pm ikz}\sin\vartheta\ket{\cal S}
\nonumber\\
&=&\left(e^{\pm ikz}+\cos^2\vartheta\,\frac{f_\pm(\theta)}{r}\,e^{ikr}\right)
\ket{M}-\cos\vartheta\sin\vartheta\,\frac{f_\pm(\theta)}{r}\,e^{ikr}\ket{m},
\end{eqnarray}
where the combination $f_\pm(\theta)=f(\theta)\pm f(\pi-\theta)$ arises 
because particles are indistinguishable and  $f(\theta)$ is the amplitude of 
scattering of flavor states. The radial part of the wave function represents a 
diverging scattered wave. One can clearly see that an initial heavy eigenstate 
acquires, upon scattering, a light eigenstate admixture. In other words,
a heavy eigenstate may be converted into a light eigenstate. This process
is illustrated in Fig.\ \ref{fig1}b. The differential cross-section of a 
process is the scattering amplitude squared. Integrating over $\theta$ we 
have the relation between the scattering and conversion cross-sections:
\begin{equation}
\sigma_{conv}=\tan^2\vartheta\,\sigma_{si}.
\end{equation}

The rest is straightforward. The light particles escape from the halo core
and decrease its density. This prevents core collapse even if the scattering
time is much smaller then the Hubble time. In this respect, the proposed
model resembles an ADM model. However, no annihilation occurs in the dense 
early Universe because forward conversions of $M$'s into $m$'s are
balanced by the reverse ones of $m$'s into $M$'s. A more detailed 
discussion of the process of freeze-out of mixed DM particles
and the resultant constraints on the model will be presented elsewhere.

\end{document}